\documentclass[letterpaper]{jpconf}
\usepackage{amsmath}
\usepackage{graphicx}

\begin{document}

\title{Early Standard Model measurements with ATLAS}
\author{Tayfun Ince, on behalf of the ATLAS collaboration}
\address{University of Victoria, Canada}
\ead{Tayfun.Ince@cern.ch}

\begin{abstract}
The measurement of Standard Model processes will be an important first
step towards exploiting the discovery potential of the Large Hadron
Collider, the highest energy accelerator ever built that will
begin operation in the fall 2009. This paper presents a summary of the
early physics analyses for understanding the performance of the
detector as well as the Standard Model at the ATLAS experiment at
14~TeV centre of mass energy.
\end{abstract}

\section{Introduction}
The Large Hadron Collider (LHC)~\cite{lhc} and the ATLAS
detector~\cite{atlas} have been designed primarily for understanding
the electroweak symmetry breaking and searching for potential new
physics beyond the Standard Model (SM). Reaching such objectives set
by the physics program places stringent requirements on understanding 
the performance of the ATLAS detector and its trigger system, the
substructure of proton and the known SM processes in the new
experimental conditions.

The analyses presented in the next sections assume 1~fb$^{-1}$ of data
or less at 14~TeV centre of mass energy, though it is likely that
there will be a period of running at 10~TeV in the first year of the
LHC operation. The LHC is capable of delivering 1~fb$^{-1}$ of data in
just a few weeks at a moderate luminosity,
$2\times10^{33}$~cm$^{-2}$s$^{-1}$. The work reported here is based on
the full simulation of the detector response to a comprehensive list
of physics processes of interest expected to be studied at the
LHC. For more details and further studies, see~\cite{cscbook}.

\section{Detector performance}
The ATLAS detector will record the signals of the decay products
of those particles of interest like the Higgs boson. In general, these
are charged leptons, photons, jets of hadrons and missing transverse
energy, $E_{{\rm T}}^{{\rm miss}}$, (e.g. neutrinos). The use of data
only techniques to measure the performance of and to calibrate the
detector is important during the initial stage of the experiment when
Monte Carlo (MC) expectations need to be re-tuned with the
understanding of this new energy regime. The well known and understood
SM processes with low background rates such as the production and
decay of the $Z$ boson, $\Upsilon$ and $J/\psi$ mesons will be used
for this purpose.

\subsection{Scale and resolution}
The design goal for the ATLAS electromagnetic calorimeters was an
overall constant term of less than 0.7\% in the energy resolution. The
local uniformity obtained from test beams and hardware calibration is
less than 0.5\% in regions of $\Delta\eta\times\Delta\phi = 0.2\times
0.4$. Inter-calibration of these regions is possible using $Z
\rightarrow ee$ events and constraining the invariant mass
distribution of electron-positron pairs to the Z boson line shape,
well-measured at LEP~\cite{lep}. The method relies on the precise
knowledge of the amount of material in front of the calorimeters. The
calibration accuracy obtained from this method is expected to be about
0.4\% using only 100~pb$^{-1}$ of data, and hence it is possible to
achieve the goal of 0.7\% global uniformity. The absolute
electromagnetic energy scale can also be obtained from this method and
is expected to be known to less than 0.5\% for the full $E_{{\rm T}}$
spectrum up to 500~GeV in the central pseudo-rapidity,
$|\eta|<1.4$, with 200~pb$^{-1}$ of data. With more statistics, the
goal is to achieve 0.1\% or better on the absolute electromagnetic
scale.

The momentum scale and the resolution of muons measured with the ATLAS
muon spectrometers can be determined using $Z \rightarrow \mu\mu$
events. An accuracy of better than 1\% in momentum scale is obtained
with 100~pb$^{-1}$ of data. The resolution for transverse momenta in
the range 10~GeV/$c$ to 500~GeV/$c$ is better than 4\% for the
spectrometer with chambers aligned to 30~$\mu$m. The resolution
degrades in the case of misaligned chambers. For instance,
misalignment of the chambers by 1~mm results in a 12\% momentum
resolution.

The ATLAS calorimeters are non-compensating, and therefore, jets
reconstructed from the calorimeter cells that are calibrated to the
electromagnetic energy scale need to be re-calibrated to the hadronic
energy scale. The accuracy of the jet calibration can be measured
using $\gamma$+jet or $Z$+jet events. One of the methods studied in ATLAS
is the $p_{{\rm T}}$ balance method in which the $p_{{\rm T}}$ of the
$\gamma$ or the $Z$ boson ($Z \rightarrow ll$ decays in particular) is
compared to that of the jet to determine the absolute hadronic energy
scale. For high $p_{{\rm T}}$ in the range 100~GeV/$c$ to 500~GeV/$c$,
$\gamma$+jet events are used, and the energy scale should be known to
a statistical precision of better than 1\% with 100~pb$^{-1}$ of
data. At low $p_{{\rm T}}$, $Z$+jet events maybe used due to lower QCD
multi-jets background rates resulting in a statistical uncertainty on
the energy scale of less than 1\% that can be achieved with
300~pb$^{-1}$ of data.

The $E_{{\rm T}}^{{\rm miss}}$ scale can be measured to better than
8\% with 100~pb$^{-1}$ of data using $Z\rightarrow \tau\tau$ events
with one $\tau$ decaying leptonically. The $E_{{\rm T}}^{{\rm miss}}$
resolution is determined using $Z\rightarrow ee$ or $\mu\mu$ events,
which should in principle have no missing energy. The resolution in
the direction longitudinal to the $Z$ boson momentum is about 3.5~GeV
when the scalar sum of the energy, $\sum E_{{\rm T}}^{{\rm cluster}}$, in
the hadronic calorimeters is 20~GeV and is better than 6~GeV for $\sum
E_{{\rm T}}^{{\rm cluster}} = 100$~GeV.

\subsection{Efficiency determination}
The trigger, reconstruction and identification efficiencies of leptons
can be determined from data only using a technique called tag \&
probe. The Drell-Yan resonances $Z$, $\Upsilon$ and $J/\psi$ will be
used to calculate the desired efficiency depending on the lepton
$p_{{\rm T}}$ range of interest. 

In the tag \& probe method, one first selects a fairly clean (low
background) sample of events with two potential lepton candidates
whose invariant mass is required to match those of one of the
resonances considered. One of these lepton candidates is required to
have passed all the selection criteria including trigger and is called
the tag. The other one is tested for a particular selection to
determine the efficiency and is called the probe. The method is found
to give consistent results with MC predictions. The systematic
uncertainty of this method is better than 2\% when the $Z$ resonance
is used. The charge misidentification rate obtained using the tag \&
probe method is less than 0.2\% in the central pseudo-rapidity region.

\section{Probing the substructure of proton}
Determining the cross sections of observed physical processes is
important when studying a new energy regime and searching for new
physics. In order to calculate a cross section at the LHC, one needs
to know the probability of finding a particular pair of partons
interacting at a momentum transfer $Q$ and carrying certain fractions,
$x$ of the proton momentum. Such information is provided in the form
of Parton Distribution Functions (PDFs), and currently these
distributions have to be determined from experimental data. Since the
LHC kinematic range in $x$ and $Q$ is much bigger than any previous
experiment, PDF uncertainty is one of the most dominant systematics,
and it can limit discovery potential for new physics. 

In ATLAS, various measurements are envisioned to further constrain
PDFs. The rapidity distribution, ${\rm d}\sigma/{\rm d}y$, of the $Z$
boson and invariant mass spectrum, ${\rm d}\sigma/{\rm d}m_{ee}$, of
low mass (20~GeV/$c^2$ - 60~GeV/$c^2$) Drell-Yan lepton pairs can be
used to constrain sea quark PDFs and low-$x$ theory in general. The
$W$ boson and $\gamma$+jet events will be used to improve gluon
PDFs. A 40\% reduction in gluon PDF uncertainty should be obtained
using 50~pb$^{-1}$ of $W\rightarrow e\nu$ data~\cite{gluonpdf}.

\section{Electroweak measurements}
Measurements of the inclusive production cross sections of the $W$ and
$Z$ bosons are crucial in understanding the SM at energies never
before studied as well as testing of precise theoretical
expectations. At 14~TeV, the production cross sections of the $W$ and
$Z$ bosons decaying into leptons are about 20~nb and 2~nb,
respectively. The measurement of the masses of the $W$ boson and the
$t$ quark at ATLAS will improve upon the already very precise values
of these parameters and will further constrain the predicted mass of
the SM Higgs boson. Further improvement on the precision of the masses
of the $W$ and $t$ will only be possible with much more than
1~fb$^{-1}$ of data.

\subsection{$W$ and $Z$ cross sections}
The typical selection of $Z\rightarrow ee$ or $\mu\mu$ events includes
using a single lepton trigger, finding a pair of oppositely charged
leptons of invariant mass in the range $m_{Z} \pm 20$~GeV/$c^2$,
requiring each lepton to have $E_{{\rm T}}>15$~GeV, $|\eta|<2.5$ and to be
isolated. The most dominant background comes from QCD multi-jet
events. The expected statistical uncertainty on the cross section is
already less than 1\% with 50~pb$^{-1}$ of data, while the systematic
uncertainty is about 4\% excluding the uncertainty on the luminosity
which is expected be better than 30\% at the early stages of data
taking.

The $W\rightarrow e\nu$ or $\mu\nu$ events are selected similarly
with a single lepton trigger, finding an isolated lepton with
$E_{{\rm T}}>25$~GeV and $|\eta|<2.5$, an $E_{{\rm T}}^{{\rm
    miss}}>25$~GeV and requiring the transverse momentum $m_T^W$ of
the $W$ be greater than 40~GeV/$c^2$. QCD multi-jets background is
also the most dominant in this case. The statistical uncertainty is
about 0.2\% with 50~pb$^{-1}$ of data. A 5\% systematic uncertainty is
expected excluding the luminosity uncertainty.

\subsection{$W$ boson and $t$ quark masses}
The measurement of the $W$ mass cannot be made directly due to
the neutrino from the $W$ decay escaping undetected. Instead, the
$p_{{\rm T}}^l$ of the lepton or the $m_{{\rm T}}^W$ of the $W$ boson,
both of which are sensitive to the mass, can be used. The $p_{{\rm
    T}}^l$ distribution has a Jacobian peak at $m_W$/2, whilst the
$m_{{\rm T}}^W$ peaks at $m_W$.

The selection of $W\rightarrow l\nu$ events include using a single
lepton trigger, requiring an isolated lepton of $p_{{\rm
    T}}>20$~GeV/$c$ and $|\eta|<2.5$, an $E_{{\rm T}}^{{\rm
    miss}}>20$~GeV and ensuring that the hadronic recoil $p_{{\rm T}}$
is less than 50~GeV/$c$. For electrons, the barrel-endcap transition
region ($1.3<|\eta|<1.6$) of the calorimeter is also excluded from
selection. In order to determine the mass of the $W$, template
distributions of $p_{{\rm T}}^l$ and $m_{{\rm T}}^W$ are generated
with various assumed values of $m_W$, smeared for detector response
and compared to the measured distribution to find the best matching
template and therefore the mass of the $W$. The $Z\rightarrow ee$ or
$\mu\mu$ events are used to measure the detector response. The
uncertainty on the mass of $W$ using $p_{{\rm T}}^l$ method with
15~pb${^{-1}}$ of data is expected to be $\Delta m_W = 120$(stat)$\pm
117$(sys)~MeV/$c^2$ in the electron channel. The most dominant
uncertainty comes from the lepton energy scale. With the $m_{{\rm
    T}}^W$ method, the uncertainty on $m_W$ is $\Delta m_W =
57$(stat)$\pm 231$(sys)~MeV/$c^2$ estimated in the muon channel. The
$E_{{\rm T}}^{{\rm miss}}$ scale is the most dominant source of
uncertainty in this method. The most dominant theoretical uncertainty
for both methods is the uncertainty on the PDFs. One important
property of $m_{{\rm T}}^W$ is its independence of the $p_T$ of the
$W$. Thus, this distribution is much less sensitive to the modelling
of the $W$ boson recoil.

The $t$ quark mass, $m_t$, will be measured at ATLAS using $t\bar{t}$
events. The production cross section of $t\bar{t}$ pairs is about
0.83~nb at 14~TeV. The most promising channel for a precise measurement is
expected to be the semi-leptonic decay in which one of the $W$ bosons
decays leptonically while the other decays hadronically resulting in a
lepton, an $E_{{\rm T}}^{{\rm miss}}$ and four-jets signature, with two of
the jets coming from the $b$ quarks. Events are selected with a single
lepton trigger and must have a lepton of $p_{{\rm T}}>20$~GeV/$c$, an
$E_{{\rm T}}^{{\rm miss}}>20$~GeV and at least four jets, three of which
with $p_{{\rm T}}>40$~GeV/$c$ and one with $p_{{\rm T}}>20$~GeV/$c$.

There are many studies performed in ATLAS to determine $m_t$. Two of
those methods are the $\chi^2$ and the geometric method. The goal of
both methods is to find those three jets (one must be a $b$-jet) which
come from the hadronically decaying $t$ quark and are to be used for
the mass determination. The $\chi^2$ method makes use of the $b$-tagging
capabilities of the detector to reduce the combinatorial background
and applies a $\chi^2$ minimization to find the two light jets which
come from the $W$ decay. The $\chi^2$ minimization constrains the mass
of the light jet pair to $m_W$ by re-scaling energies of the jets. The
light and $b$-jet energy scales are the most dominant systematic
uncertainties for the $\chi^2$ method. The effect of these scales on
the $t$ quark mass uncertainty is 0.2~GeV/$c^2$/\% and 0.7~GeV/$c^2$/\%
respectively. For example, a 1\% $b$-jet energy scale uncertainty
gives an expectation of a 0.7~GeV/$c^2$ uncertainty on the mass of the
$t$ quark. Assuming both the light and $b$-jet energy scale
uncertainties are known to 1\%, the total systematic uncertainty on
$m_t$ with this method is about 1~GeV/$c^2$. The statistical
uncertainty is less than 0.4~GeV/$c^2$ with 1~fb$^{-1}$ of data. The
geometric method, on the other hand, does not use $b$-tagging and is
complementary to the $\chi^2$ method especially during the early
stages of the experiment until the detector calibration matures. This
method simply chooses the two geometrically closest jets with
invariant mass around the $m_W$. Lack of b-tagging in the geometric
method results in a higher rate of combinatorial background and an
uncertainty of about 1~GeV/$c^2$ on $m_t$. The overall systematic
uncertainty on $m_t$ with this method is 2~GeV/$c^2$ assuming 1\%
uncertainty on the jet (light or $b$) energy scale.
  
\section{Conclusions}
Studying SM processes and re-measuring its parameters are not only
important for understanding of the ATLAS detector, but are also essential
in order to be able to discover new physics. The statistical
uncertainties on SM measurements will be negligible at LHC. Therefore
it is possible to improve the uncertainty even on the already
precisely measured SM parameters as understanding of the response of
the ATLAS detector advances.

\end{document}